\documentclass[prb,twocolumn]{revtex4}
\usepackage{amsmath}
\usepackage{amssymb}
\usepackage{graphicx}

\begin{document}

\title{Conductance of a quantum point contact based on
spin-density-functional theory}
\author{S. Ihnatsenka}
\affiliation{Solid State Electronics, Department of Science and Technology (ITN), Link%
\"{o}ping University, 60174 Norrk\"{o}ping, Sweden}
\author{I. V. Zozoulenko}
\affiliation{Solid State Electronics, Department of Science and Technology (ITN), Link%
\"{o}ping University, 60174 Norrk\"{o}ping, Sweden}
\date{\today}

\begin{abstract}
We present full quantum mechanical conductance calculations of a quantum
point contact (QPC) performed in the framework of the density functional
theory (DFT) in the local spin-density approximation (LDA). We start from a
lithographical layout of the device and the whole structure, including
semi-infinitive leads, is treated on the same footing (i.e. the
electron-electron interaction is accounted for in both the leads as well as
in the central device region). We show that a spin-degeneracy of the
conductance channels is lifted and the total conductance exhibits a broad
plateau-like feature at $\sim 0.5\times 2e^{2}/h$. The lifting of the
spin-degeneracy is a generic feature of all studied QPC structures (both
very short and very long ones; with the lengths in the range $40\lesssim
l\lesssim 500$ nm). The calculated conductance also shows a hysteresis for
forward- and backward sweeps of the gate voltage. These features in the
conductance can be traced to the formation of weakly coupled quasi-bound
states (magnetic impurities) inside the QPC (also predicted in previous
DFT-based studies). A comparison of obtained results with the experimental
data shows however, that while the spin-DFT based "first-principle"
calculations exhibits the spin polarization in the QPC, \textit{the
calculated conductance clearly does not reproduce the 0.7 anomaly observed
in almost all QPCs of various geometries.} We critically examine major
features of the standard DFT-based approach to the conductance calculations
and argue that its inability to reproduce the 0.7 anomaly might be related
to the infamous derivative discontinuity problem of the DFT leading to
spurious self-interaction errors not corrected in the standard LDA. Our
results indicate that the formation of the magnetic impurities in the QPC
might be an artefact of the LDA when localization of charge is expected to
occur. We thus argue that an accurate description of the QPC structure would
require approaches that go beyond the standard DFT+LDA schemes.
\end{abstract}

\maketitle

\section{Introduction}

Experimental evidence of an additional $0.7\times 2e^{2}/h$
conductance feature in quantum point contacts
(QPCs)\cite{Thomas1996, Thomas1998, trench-etchedQPC, epitaxQPC,
Reillytwin, Cronenwett, Reilly_2002,
Reilly_2005,Graham,Rokhinson}  has generated enormous theoretical
activity during recent decade\cite%
{Cronenwett, Reilly_2002, Reilly_2005, Berggren, Starikov, Jaksch,
Havu, Meir_2002, Meir_2003, Meir_Nature, Shelykh}. While a usual
conductance quantization in terms of $2e^{2}/h$ can be
successfully explained in a one-electron picture\cite{Wharam}, no
consensus on the origin of the 0.7 anomaly has been reached so
far. Experimental data, based on magnetic field dependence of the
plateau position\cite{Thomas1996,Thomas1998} and observation of
the zero bias anomaly\cite{Cronenwett} clearly point out at spin
origin of this effect. The spin origin was adopted in many
theories which include, just to name a few of them, a spontaneous
spin-polarization inside the constriction of the QPC originated
from the exchange-correlation interaction\cite{Berggren, Starikov,
Jaksch, Havu}, the ferromagnetic-antiferromagnetic exchange
interaction with a large localized spin\cite{Shelykh}, a formation
of polarized quasi-bound states and Kondo effect\cite{Meir_2002,
Meir_2003}, as well as several others. However, none of the
existing theories reproduce \textit{quantitatively} the 0.7
conductance anomaly observed in almost all QPCs\cite{Thomas1996,
Thomas1998, trench-etchedQPC, epitaxQPC, Reillytwin, Cronenwett,
Reilly_2002, Reilly_2005}. Note that some of approaches based on
phenomenological models
do recover the 0.7 feature in the conductance\cite%
{Reilly_2002,Reilly_2005,Bruus}. Such the approaches, while providing an
important insight for an interpretation of the experiment, are not, however,
able to uncover the microscopic origin on the observed effect.

A powerful technique capable of providing detailed information about
electronic and transport properties on the microscopic level is based on the
"first principle" density functional theory (DFT) approach. The spin DFT
calculations addressing the 0.7 anomaly in the QPC have been presented by
various groups\cite{Berggren,Starikov,Jaksch,Havu,Meir_2002,Meir_Nature}.
However, many of these calculations show rather conflicting results. For
example. Refs. \onlinecite{Meir_2002,Meir_2003,Meir_Nature} attribute the
0.7 anomaly to formation of the localized magnetic moment in the QPC,
whereas such quasi-bound states are not recovered in Refs. %
\onlinecite{Berggren, Starikov, Jaksch,Havu}. In contrast, Ref. %
\onlinecite{Starikov} relates the $0.7$ anomaly to multiple metastable
spin-polarized solutions. The recent study of Jaksch \textit{et al.}\cite%
{Jaksch} suggests that the spin polarization is absent in short QPCs and is
increased with the increase of the length of the QCP. This seems to
contradict the results reported by Rejec and Meir\cite{Meir_Nature} who find
that a strong spin polarization due to the magnetic impurity formation is a
generic feature seen in short as well as in long QPCs.

Motivated by the above mentioned conflicting results and conclusions, in
this paper we perform \textquotedblleft first principle\textquotedblright\
full quantum mechanical transport calculations of the conductance of the QPC
based on the spin DFT in the local spin-density approximation (LDA). [We use
a notation \textquotedblleft first principle\textquotedblright\ in a sense
that we start from a lithographical geometry of the gates and a
heterostructure layout]. The main features of our calculations and the
relation of our model to the DFT-based calculations reported previously are
as follow. In the present paper the QPC is viewed as an inherently \textit{%
open} system where the solution of the Schr\"{o}dinger equation consists of
continuous scattering states. This is in contrast to Refs. %
\onlinecite{Berggren, Jaksch} treating a QPC as a closed system where the
solution of the Schrodinger equation is represented by a discrete set of the
eigenfunctions. In our calculations we start from a lithographical layout of
the device, which is in contrast to Ref. \onlinecite{Havu} using a
simplified model of an external confinement potential. (Note that importance
of an accurate treatment of the confinement potential is stressed in Ref. %
\onlinecite{Reilly_2005}, where the phenomenology and experimental data
indicate that the 0.7 feature depends strongly on the potential profile of
the contact region). Our model (but not the computational technique) is
conceptually similar to the one used by Rejec and Meir\cite{Meir_Nature}
(see also Ref. \onlinecite{Meir_2002} reporting similar findings) where a
QPC is treated as an open structure and where a realistic model for the
external confinement potential due to metallic gates is utilized. In their
study Rejec and Meir focus on the calculation of the electron density,
predicting the formation of a localized spin-degenerate quasi-bound state
(magnetic impurity). They did not, however, perform transport conductance
calculations so the central question whether the 0.7-anomaly is indeed
related to the formation of a magnetic impurity has remained unanswered. In
the present paper we hence focus on the calculation of the spin-resolved
conductance of the quantum point contact.

The details of our model and the Hamiltonian are given in Sec. II. In Sec.
III we present our method for the calculation of the electron density and
conductance which is based on the self-consistent Greens function technique
combined with the spin DFT in the local spin density approximation. This
method is a generalization of a method of Ref. \onlinecite{opendot} for the
case of spin. The results are presented in Sec. IV. Our calculations
reconfirm the finding reported by Rejec and Meir\cite{Meir_Nature}
concerning the formation of the localized spin-degenerate quasi-bound states
(magnetic impurities) within the DFT+LDA approach. However, \textit{the
calculated conductance clearly does not reproduce the 0.7 anomaly observed
in almost all QPCs of various geometries.} Instead, the total conductance
shows a broad feature peaked at $0.5\times 2e^{2}/h$. A similar feature is
also present in the range of the gate voltages where a second step in the
conductance develops. The calculated conductance also shows a hysteresis for
forward- and backward sweeps of the magnetic fields. (We stress that all
these results are generic; we studied QPS with lengths in the range 40 - 500
nm and electron densities in the leads in the range $10^{15}\text{m}%
^{-2}-4\cdot 10^{15}\text{m}^{-2}$, with very similar results). In Sec. V we
discuss the obtained results and critically examine the DFT+LDA-based
conductance calculations. We suggest that the failure of the DFT+LDA
approach to reproduce quantitatively the 0.7 anomaly may be due to the lack
of the derivative discontinuity in the standard LDA approximation (leading
to uncorrected self-interaction errors) for the case when localization of
charge is expected to occur, so that the magnetic impurity formation may be
an artefact of the DFT+LDA approach due to the spurious self-interaction.

\section{Model}

\begin{figure}[tb]
\includegraphics[keepaspectratio,width=\columnwidth]{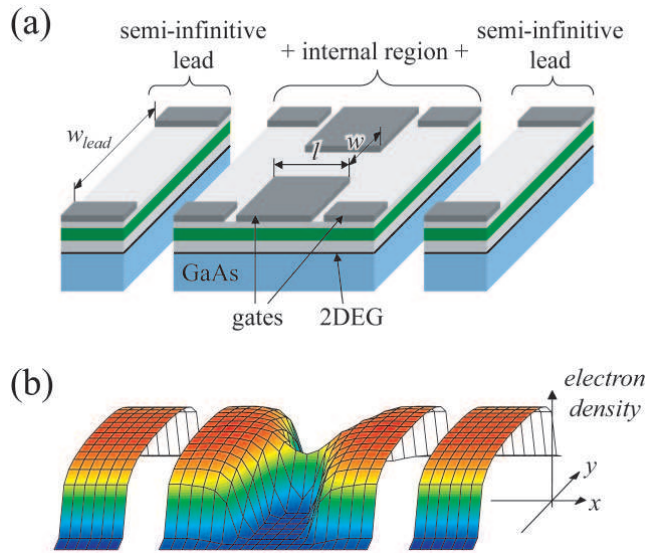} 
\caption{(color online) (a) Structure of a quantum point contact
typically fabricated by a split-gate technique. An internal
computational region is attached to two semi-infinite leads which
serve as electron reservoirs. The gates of the quantum point
contact form a constriction of the length $l$ and the width $w$;
$w_{lead}$ is the distance between the gates defining the leads.
(b) A representative calculated electron density.}
\label{f:structure}
\end{figure}

We consider a quantum point contact (QPC) placed between two semi-infinite
electron reservoirs. A schematic layout of the device is illustrated in Fig. %
\ref{f:structure}(a). Charge carriers originating from a fully ionized donor
layer form the two-dimensional electron gas (2DEG) which is buried inside a
substrate at the GaAs/Al$_{x}$Ga$_{1-x}$As heterointerface at the distance $%
b $ from the surface. Metallic gates are situated on the top of the
heterostructure and define the quantum point contact as well as electron
reservoirs that are are represented by uniform quantum wires of the infinite
length.

The Hamiltonian of the whole system (QPC + leads) within the Kohn-Sham
formalism can be written in the form $H=\sum_{\sigma }H^{\sigma },$
\begin{equation}
H^{\sigma }=-\frac{\hbar ^{2}}{2m^{\ast }}\nabla ^{2}+V^{\sigma }(\mathbf{r}%
),  \label{Hamiltonian}
\end{equation}%
$\mathbf{r}=(x,y)$, $m^{\ast }=0.067m_{e}$ is the GaAs effective mass and $%
\sigma $ stands for spin-up, $\uparrow $, and spin-down, $\downarrow $,
electrons. The first term in (\ref{Hamiltonian}) is the kinetic energy of an
electron while $V^{\sigma }(\mathbf{r})=V_{conf}(\mathbf{r})+V_{H}(\mathbf{r}%
)+V_{xc}^{\sigma }(\mathbf{r})$ is the total confining potential which is a
sum of the electrostatic confinement potential, the Hartree potential, and
the exchange-correlation potential, respectively. The electrostatic
confinement $V_{conf}(\mathbf{r})=V_{gates}(\mathbf{r}%
)+V_{donors}+V_{Schottky}$ includes contributions from the top gates, the
donor layer and the Schottky barrier. The explicit expressions for the
potentials $V_{gates}(\mathbf{r})$ and $V_{donors}$ are given in Refs. %
\onlinecite{Davies_gate} and \onlinecite{Martorell_donor}; the Schottky
barrier is chosen to be $V_{Schottky}=0.8$ eV. The Hartree potential is
written in a standard form%
\begin{equation}
V_{H}(\mathbf{r})=\frac{e^{2}}{4\pi \varepsilon _{0}\varepsilon _{r}}\int d%
\mathbf{r}\,^{\prime }n(\mathbf{r}^{\prime })\left( \frac{1}{|\mathbf{r}-%
\mathbf{r}^{\prime }|}-\frac{1}{\sqrt{|\mathbf{r}-\mathbf{r}^{\prime
}|^{2}+4b^{2}}}\right) ,  \label{V_H}
\end{equation}%
where $n(\mathbf{r})=\sum_{\sigma }n^{\sigma }(\mathbf{r})$ is the total
electron density and the second term describes the mirror charges placed at
the distance $b$ from the surface, $\varepsilon _{r}=12.9$ is the dielectric
constant of GaAs, and the integration is performed over the whole device
area including the semi-infinite leads.

The exchange-correlation potential $V_{xc}^{\sigma }(\mathbf{r})$ in the
local spin density approximation is given by the functional derivative \cite%
{Giuliani_Vignale}
\begin{equation}
V_{xc}^{\sigma }=\frac{d}{dn^{\sigma }}\left\{ n\epsilon _{xc}\left(
n\right) \right\} .  \label{LDA}
\end{equation}%
For $\epsilon _{xc}$ we have employed two commonly used parameterizations,
by Tanatar and Ceperley\cite{TC} and by Attacalite \textit{et al}\cite{AMGB}%
. These two parameterizations give very similar results. All the results
presented below correspond to the parameterization of Tanatar and Ceperley%
\cite{TC}.

\section{Method}

The central quantity in transport calculations is the conductance. In the
linear response regime, it is given by the Landauer formula $G=\sum_{\sigma
}G^{\sigma }$, \cite{Datta_book}
\begin{equation}
G^{\sigma }=-\frac{e^{2}}{h}\int dE\,T^{\sigma }(E)\frac{\partial f\left(
E-E_{F}\right) }{\partial E},  \label{conductance}
\end{equation}%
where $T^{\sigma }(E)$ is the total transmission coefficient for the spin
channel $\sigma $, $f(E-E_{F})$ is the Fermi-Dirac distribution function and
$E_{F}$ is the Fermi energy. In order to calculate $T^{\sigma }(E)$ we use
the method developed in Ref. \onlinecite{opendot} for the spinless electrons
and generalize it here for the presence of two spin channels. For the case
of completeness, the main steps of our calculations are presented below.

We discretize Eq. (\ref{Hamiltonian}) and introduce the tight-binding
Hamiltonian with the lattice constant of $a=4$ nm. (Such a small $a$ ensures
that the tight-binding Hamiltonian is equivalent to the continuous Schr\"{o}%
dinger equation.) The retarded Green's function is introduced in a standard
way\cite{Datta_book},
\begin{equation}
\mathcal{G}^{\sigma }=\left( E-H^{\sigma }+i\eta \right) ^{-1}.
\label{gfunction}
\end{equation}%
The Green's function in the real space representation, $\mathcal{G}^{\sigma
}(\mathbf{r},\mathbf{r},E)$, provides an information about the electron
density at the site $\mathbf{r},$ \cite{Datta_book}
\begin{equation}
n^{\sigma }(\mathbf{r})=-\frac{1}{\pi }\Im \int dE\,\mathcal{G}^{\sigma }(%
\mathbf{r},\mathbf{r},E)\,f(E-E_{F}).  \label{density}
\end{equation}%
Note that $\mathcal{G}^{\sigma }(\mathbf{r},\mathbf{r},E)$ is a rapidly
varying function of energy. As a result, a direct integration along the real
axis in Eq. (\ref{density}) is rather ineffective as its numerical accuracy
is not sufficient to achieve a convergence of the self-consistent electron
density. Because of this, we transform the integration contour into the
complex plane $\Im \lbrack E]>0,$ where the Green's function is much more
smoother (see Ref. \onlinecite{opendot} for details).

In order to calculate the Green's function of the whole system (QPC + leads)
we divide it into three parts, the internal region and two semi-infinite
leads, as shown in Fig. \ref{f:structure}(a). The internal region consists
of the QPC as well as a part of the leads. In order to link the internal
region and the leads together the self-consistent charge density (and the
potential) at both sides must be the same. To fulfill this requirement, we
place the semi-infinite leads far away from the central region so that the
gates defining the QPC do not affect the electron density distribution in
the leads. (The distribution is determined solely by the lead gate voltage,
see Fig. \ref{f:structure}.) As a result, it is fully justified to
approximate the semi-infinitive leads by a uniform quantum wire. The
self-consistent solution for the latter can be found by the technique
developed in Ref. \onlinecite{Ihnatsenka}. Eventually, the Green's function
of the whole system is calculated by linking the surface Green's function
for the semi-infinite leads (calculated using the technique of Ref. %
\onlinecite{Ihnatsenka}) and the Green's function of the internal region
with the help of the the Dyson equation\cite{Zozoulenko_1996}.

All the calculations described above are performed self-consistently in an
iterative way until a converged solution for the electron density and
potential (and hence for the total Green's function) is obtained. Having
calculated the total self-consistent Green's function, the scattering
problem is solved where the scattering states in the leads (both propagating
and evanescent) are obtained using the Green's function technique of Ref. %
\onlinecite{Ihnatsenka}.

Having calculated the Green's function we can find the local density of
states (LDOS) in the QPC\cite{Datta_book}
\begin{equation}
\text{LDOS}^{\sigma }(x,E)=-\frac{1}{\pi }\Im \int dy\;\mathcal{G}^{\sigma
}(x,y,E).  \label{dos}
\end{equation}

The self-consistent solution in quantum transport or electronic structure
calculations is often found using a \textquotedblleft simple
mixing\textquotedblright\ method. It is the most robust and reliable
algorithm with only one disadvantage, namely a low convergence rate. The
charge density on the ( $i+1$)-th iteration loop is updated through the
input $n_{i}^{in}(\mathbf{r})$ and output $n_{i}^{out}(\mathbf{r})$
densities on the previous $i$-th iteration
\begin{equation}
n_{i+1}^{in}(\mathbf{r})=(1-\epsilon )n_{i}^{in}(\mathbf{r})+\epsilon
\,n_{i}^{out}(\mathbf{r}),  \label{iteration}
\end{equation}%
with $\epsilon $ being a small constant $\sim 0.1-0.01$. It is typically
needed $\sim 200-2000$ iteration steps to achieve our convergence criterium
for the relative density update on the $i$-th iteration step,
\begin{equation}
\frac{|n_{i}^{out}-n_{i}^{in}|}{n_{i}^{out}+n_{i}^{in}}<10^{-7},
\label{criteria}
\end{equation}%
where $n=\int n(\mathbf{r})\,d\mathbf{r}$ is the total electron number in
the internal region during the $i$-th iteration step.

In order to trigger a spin-polarized solution we apply at early iterations a
small parallel magnetic field, $B=0.05$ T. The final solution is driven
mostly by the exchange interaction which is about four orders of magnitude
larger that the Zeeman energy at $B=0.05$ T.

The self-consistent calculation are performed for different gate voltages.
To facilitate the calculations we use a solution from the previous value of
the gate voltage as an initial guess for the subsequent one. It is worth
noting that the modified Broyden's second method\cite{Singh}, which can
greatly reduce the number of iterations for the case of spinless electrons%
\cite{opendot}, does not lead to reliable convergent results for the case of
the spin degree of freedom.

\section{Results}

We calculate the conductance of a split-gate QPC with following parameters
representative for a typical experimental structure. The 2DEG is buried at $%
b=70$ nm below the surface (the widths of the cap, donor and spacer layers
are 24 nm, 36 nm and 10 nm respectively), the donor concentration is $%
0.6\cdot 10^{24}$ m$^{-3}$. The width of the semi-infinitive leads is $%
w_{lead}=400$ nm, and the width of the QPC constriction is $w=100$ nm, see
Fig. \ref{f:structure} (a). The width $w$ is kept constant throughout the
paper, while the length is varied in the range $l=40-500$ nm. The gate
voltage applied to the gates defining the leads is $V_{lead}=-0.73$ V. With
these parameters of the device there are 18 spin-up and 18 spin-down
channels available for propagation in the leads and the electron density in
the center of the leads is $n_{lead}^{\uparrow }=n_{lead}^{\downarrow
}=2.2\cdot 10^{15}$ m$^{-2}$. Such a large number of the channels in the
leads makes its DOS to mimic a smooth DOS of the two-dimensional electron
gas. The temperature is fixed at $T=0.2$ K for all results presented below.

\begin{figure}[tb]
\includegraphics[scale=1.0]{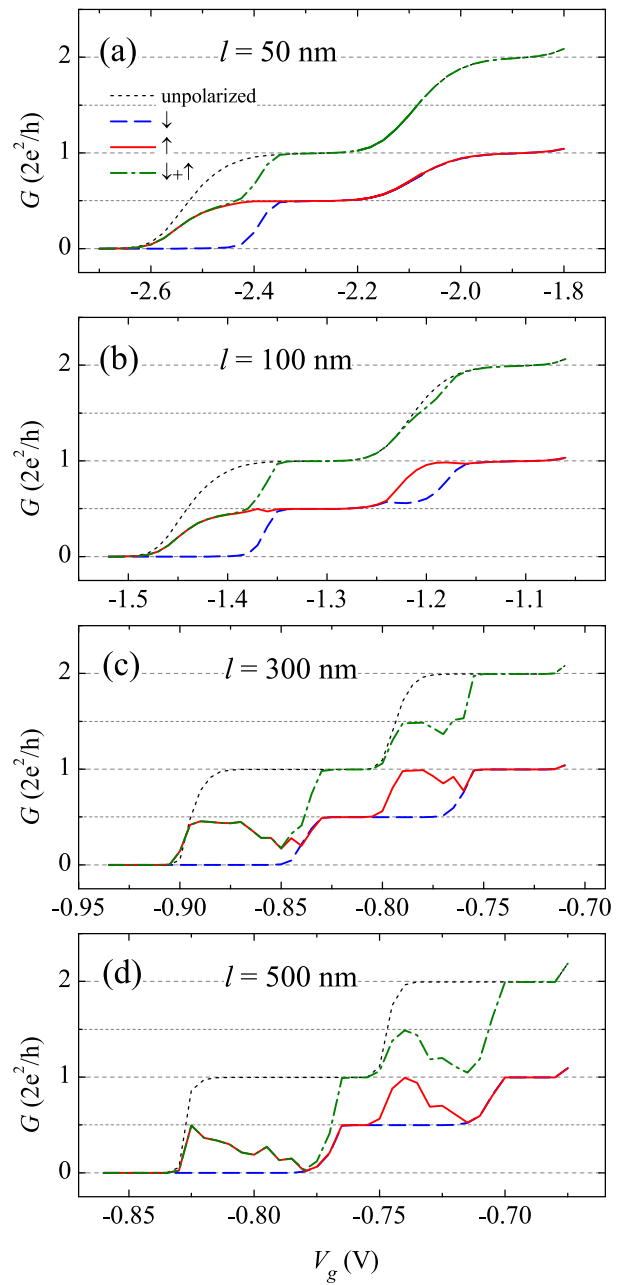}
\caption{(color online) Conductance of the quantum point contact
of different lengths $50 < l < 500$ nm as a
function of the gate voltage $V_g$. The geometrical width of the constriction is $%
w=100$ nm; the geometrical length $l$ is indicated in the figures. The
dashed line corresponds to the spin-unpolarized solution.}
\label{f:G}
\end{figure}

Figure \ref{f:G} shows the conductance of QPCs of different
constriction lengths $l$ ranging from $l=50$ nm (a very short QPC)
to $l=500$ nm (a long quantum wire-type QPC). The conductance of
all QPCs shows a broad plateau-like feature at $0.5\times
2e^{2}/h$. As the length of the constriction increases, a dip
following 0.5-plateau starts to develop in the QPC conductance. An
inspection of the spin-resolved conductance demonstrates that 0.5
feature corresponds to the transmission of only one spin channel
(say, spin-up), whereas the second (spin-down) conductance channel
is totally suppressed. For long constrictions ($l\gtrsim 300$ nm)
the 0.5 plateau starts to \textquotedblleft wear
down\textquotedblright\
transforming into a broad feature whose maximal amplitude is less than $%
0.5\times 2e^{2}/h$. If the constriction is sufficiently long ($l\gtrsim 100$
nm), a conductance plateau at $\sim 1.5\times 2e^{2}/h$ starts to develop
and a conductance dip following the 1.5-plateau also starts to emerge as the
length of the constriction increases.

To shed light on a microscopic origin of the $0.5\times 2e^{2}/h$
conductance feature and the suppression of the spin-down channel
let us inspect the charge density and energy structure of the QPC.
\begin{figure}[tb]
\includegraphics[keepaspectratio,width=\columnwidth]{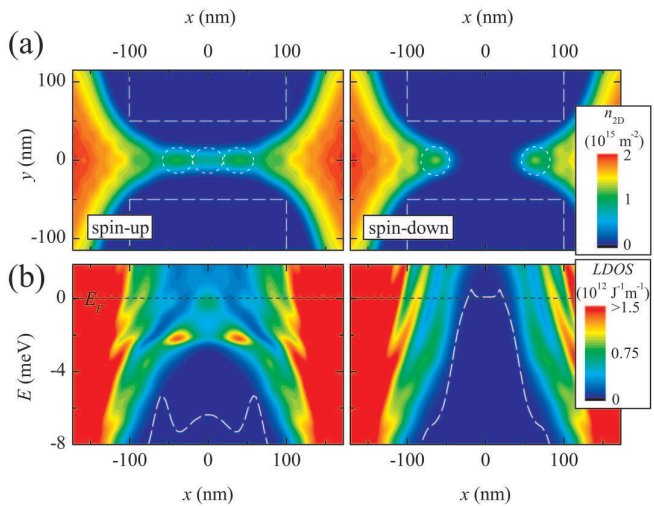}
\caption{(color online) Formation of quasi-bound states in the quantum point
contact. (a) The charge density and (b) the local density of states are
shown for the regime of one transmitted spin-up and totally blocked
spin-down channel. The left and right columns correspond to spin-up and spin
down electrons. The white dashed lines in (b) indicate the self-consistent
Kohn-Sham potential along the cross section $y=0$, $V^{\protect\sigma %
}(x,y=0)$. The charge droplets with approximately one electron trapped are
denoted by dotted lines in (a). The geometrical length and width of the QPC
are $l=200$ nm and $w=100$ nm, respectively; $V_{g}=-0.98$ V.}
\label{f:droplet}
\end{figure}
Figure \ref{f:droplet} shows the charge densities and the LDOS in the QPC of
the length of $l=200$ nm for the case of one transmitted spin-up, $%
G^{\uparrow }=e^{2}/h$, and totally blocked spin-down, $G^{\downarrow }=0$,
channel. Near the constriction entrance, the spin-polarized charge droplets
(marked by dotted circles in Fig. \ref{f:droplet}(a)) are clearly visible.
The spin polarization is caused by the exchange interaction which dominates
the kinetic energy for low densities. The integration of the electron
density gives about one particle in each droplet. The quasi-bound states for
corresponding droplets can also be traced in the LDOS shown in Fig. \ref%
{f:droplet}(b). An inspection of the corresponding potential profile for the
spin-up electrons reveals that the low- and high energy droplets correspond
to respectively first and second quasibound states trapped in the
double-barrier potential. At the same time the spin-down droplets are
spatially separated by a distance $\sim 100$ nm. This spatial separation is
also reflected in the shape of the total confining potential for the
spin-down electrons that forms a wide tunnelling barrier, see Fig. \ref%
{f:droplet} (b). Because of this barrier, the transmission probability for
the spin-down conductance is negligibly small ($G^{\downarrow }\approx
10^{-5}\times e^{2}/h$).

The quasi-bound states are not spatially fixed but gradually move during the
sweep of the gate voltage, $V_{g}$. Already in the pinch-off regime, two
quasi-bound states are developed at both sides of the constriction. When $%
V_{g}$ becomes less negative they move towards each other and eventually
merge. Because of the exchange interaction, this occurs first for the
spin-up state and then for the spin-down. As a result, the total conductance
$G$ shows quantization in units of $e^{2}/h$, see Fig. \ref{f:G}. We stress
that all the results reported above are generic; we studied QPCs with
lengths in the range 40 - 500 nm and electron densities in the leads in the
range $10^{15}\text{m}^{-2}-4\cdot 10^{15}\text{m}^{-2}$, with very similar
results. We also stress that our calculations with two parameterizations of
Refs. \onlinecite{TC,AMGB} for the correlation potential give almost the
same results. This is not surprising, because for the system at hand the
correlation potential is an order of magnitude lower than the exchange
potential.

Our conclusions concerning formation of the localized quasi-bound states in
the QPC agree well with earlier findings of Hirose \textit{et al.}\cite%
{Meir_2002} and Rejec and Meir\cite{Meir_Nature}, but do not support the
conclusion of Jaksch \textit{et al. }\cite{Jaksch} that the spin
polarization is absent in short QPCs and is increased with the increase of
the length of the QCP.

\begin{figure}[tb]
\includegraphics[keepaspectratio,width=\columnwidth]{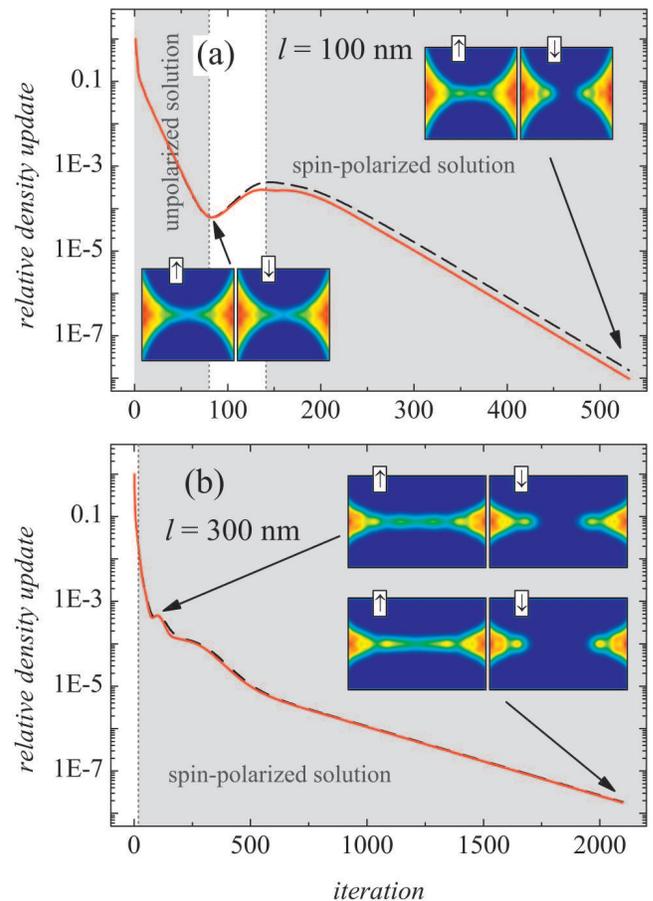} 
\caption{(color online) Convergence of self-consistent calculations for (a)
a short QPC ($l=100$ nm) and (b) a long QPC ($l=300$ nm) ($y$-axis
corresponds to the relative density update defined by Eq. (\protect\ref%
{criteria})). Solid and dashed lines denote spin-up and spin-down
solutions, respectively. Insets show the charge densities at the
constriction at different stages of iteration procedure. Note,
that we used the convergence criteria for the relative density
update $< 10^{-7}$ for all calculations presented in this paper.
The geometrical width of the constriction $w=100$ nm; $V_{g}=-1.4$
V.} \label{f:convergence}
\end{figure}

We now want to focus on an aspects of the calculations concerning a choice
of the criteria for termination of the iteration procedure, Eq. (\ref%
{criteria}). Even though this aspect has a rather technical character, we
feel that it is important to address it in detail, as the proper choice of
the termination criteria is essential for reaching of the spin polarized
solution. Figure \ref{f:convergence} shows a relative density update as a
function of the number of iterations for a relatively short ($l=100$ nm) and
a relatively long ($l=300$ nm) constrictions. For the long constriction the
spin polarized solution is obtained already during initial iteration steps
and its character does not change with increase of the number of iterations,
see Fig. \ref{f:convergence} (b). On the contrary, for a short constriction,
the spin polarized solution is obtained only during later stages of
iterations, as no spin-polarized solution is present during the initial
steps even when the relative density update decreases to $\sim 10^{-4},$ see %
\ref{f:convergence} (a). Note, that the spin-polarized solution is
energetically favorable and more stable than the spin-unpolarized one for
all QPC studied. %
\begin{figure}[tb]
\includegraphics[keepaspectratio,width=\columnwidth]{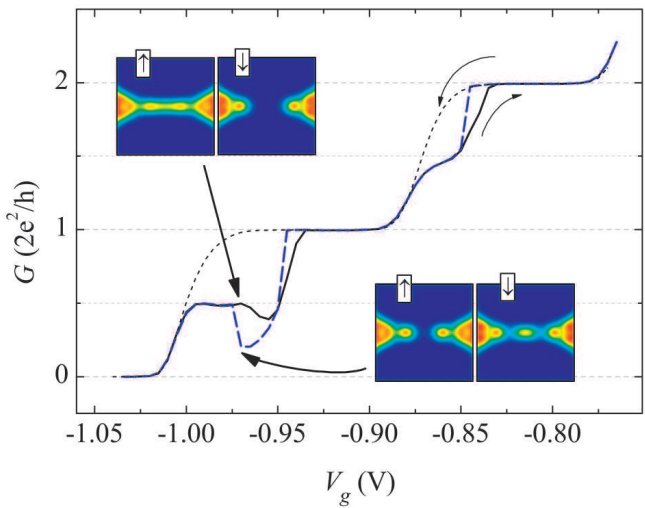} 
\caption{(color online) Conductance hysteresis in the quantum point contact
for forward- and backward sweeps of the gate voltage (solid and dashed lines
respectively). The dotted line corresponds to the unpolarized solution
exhibiting no hysteresis behavior. Insets show the charge densities in the
constriction at $V_{g}=-0.975$ V. The parameters of QPC are $l=200$ nm and $%
w=100$ nm.}
\label{f:hysteresis}
\end{figure}

The spin-DFT calculations reported above show the presence of the
spin-resolved quasi-bound states confined in an effective double-barrier
potential. It is well known that the quasi-bound states often lead to the
hysteretic behavior of the system. For example, the hysteresis is commonly
observed in the \textit{I-V} characteristics of the resonant double-barrier
tunnelling structures for forward- and backward voltage sweeps\cite%
{Goldman,RTD-book}. Such hysteretic behavior is related to the presence of
the quasi-bound state between the barriers and was reproduced in numerous
calculations\cite{RTD-book}, including those based on the DFT approach\cite%
{RTDbistable}. Our calculations demonstrate that similar hysteretic behavior
is present in the conductance of the QPC. When the gate voltage is swept
forward and backwards, the conductance shows the prominent hysteresis in the
transition regions between the plateaus, see Fig. \ref{f:hysteresis}. In the
region of hysteresis, the system, depending on the history, can be in one of
two different ground states as illustrated in the inset to Fig. \ref%
{f:hysteresis}. Because the hysteresis is present only in the transition
regions and is completely absent in the spin-unpolarized solution, its
origin is due to the exchange interaction. We would like to stress that
existence of two distinct ground states as well as the hysteresis behavior
of the system at hand does not contradict to the DFT approach as such. Owing
to the nonlinearity of the problem, there might be more than one solution
within the DFT framework\cite{Almbladh}. Besides the double-barrier resonant
tunnelling structures mentioned above\cite{RTDbistable}, the DFT approach
was used for the description of the hysteretic behavior of quantum Hall
states in the double-layered quantum well structures\cite{MacDonald}, as
well as hysteresis and spin phase transitions in quantum wires in the
integer quantum Hall regime\cite{hysteresis}. We however, conclude this
section by noticing that we are not aware of any experimental reports on the
observation of the hysteresis phenomena in the QPC structures.

\section{Discussion}

We start our discussion from a comparison of the spin DFT-based conductance
calculations reported in the preceding section with the experimental
conductance of the QPC structures. The calculations show a pronounced
plateau-like feature at $\sim 0.5\times 2e^{2}/h$ for all QPC structures
studied (both short and long ones). For longer QPCs ($l\gtrsim 150$ nm), a
dip following 0.5-plateau starts to develop in the conductance. At the same
time, a new conductance plateau at $\sim 1.5\times 2e^{2}/h$ starts to
develop and a conductance dip following 1.5-plateau also starts to emerge as
the length of the constriction increases (see Fig. \ref{f:G}). Our
calculations also show a pronounced hysteretic behavior in the transition
regions between the plateaus for forward and backward sweeps of the gate
voltage, Fig. \ref{f:hysteresis}. On the contrary, the experimental data
clearly show an anomaly in the conductance around $0.7\times 2e^{2}/h.$ [It
should be stressed that the calculated 0.5 plateau corresponds to the
complete suppression of one spin channel, whereas experimentally observed
0.7 anomaly means that both spin channels have to be present in the
conductance]. Some experiments\cite{Reilly_2005} also indicate a feature
around $1.7\times 2e^{2}/h$. As far as the hysteresis behavior is concerned,
we are not aware of any experimental reports of this effect. \textit{The
above comparison demonstrates that while the spin-DFT based
\textquotedblleft first-principle\textquotedblright\ calculations predict
the spin polarization in the QPC structure, the calculated conductance
clearly does not reproduce the 0.7 anomaly observed in almost all QPCs of
various geometries.}

In order to understand why the calculated conductance fails to reproduce the
0.7 anomaly, let us critically analyze the major features of the DFT-based
conductance calculations. Our computation scheme relies on an approach that
during recent years became a standard tool for transport calculations in
open electronic systems such as molecules, metallic wires and mesoscopic
conductors\cite{Lang,Guo_open_dot,Guo,Ratner,Datta,Brandbyge}. Its starting
point is the Landauer-type formula where the conductance is calculated using
the Green's function or similar scattering techniques combined with the
mean-field description of the electronic structure in the leads and in the
device region typically based on the ground state self-consistent Kohn-Sham
orbitals. Because of the conceptual appeal of this approach, the earlier
works were not focused on its formal justification. Recently, however,
several studies have provided the rigorous theoretical foundation for the
above DFT-based methodologies calculating the steady currents in open
electronic systems\cite{Almbladh,Evers,Chen,Ferretti}. Therefore, the
question concerning the validity of the present approach relies mostly on
the proper description of the exchange and correlation within the DFT
approximation. We focus below only on two aspects of the DFT approach that
seem to be the most important for understanding of the discrepancy between
the calculations and the experiment.

The exchange and correlations are commonly accounted for within the local
spin density appromaximation\cite{Giuliani_Vignale}. As mentioned in
preceding sections, for two dimensional systems there are two commonly used
parameterizations, namely the parameterization of Tanatar and Ceperley\cite%
{TC} and Attacalite \textit{et al}\cite{AMGB}. The validity of these
approximations was tested for few-electron quantum dot systems and,
generally, a very good agreement agreement with the exact diagonalization
and/or variational Monte Carlo calculations is found\cite{validity}. Taking
into account that these parametrizations give practically the same results
for the QPC conductance, we do not expect the utilization of the above
parametrizations to be a source of a  significant discrepancy between the
calculated conductance and the experiment. Instead, we focus on another
aspect of the choice of the exchange energy functional which arises in the
systems with a variable particle number. (Note that the QPC structure, being
an essentially open structure, belongs to this class of systems). This
aspect is related to the infamous \textquotedblleft derivative discontinuity
problem\textquotedblright\ of the DFT originating from the discontinuous
dependence of $V_{xc}$ on the particle number\cite{Giuliani_Vignale}. (Note
that the LDA does not include any derivative discontinuity in the $V_{xc}$).

The problem with the derivative discontinuity (leading to uncompensated
self-interaction corrections in the $V_{xc}$) has been recently recognized
in the standard quantum mechanical transport calculations in molecular
systems and atomic wires\cite{Evers,Toher,Koentopp}. Typically, such the
calculations provide a good quantitative agreement with the experimental
data for systems where the coupling to the leads is strong and the
conductance exceeds the conductance unit $G_{0}=2e^{2}/h$ (for example, for
atomistic metallic wires and related systems). At the same time, for weakly
coupled systems such as organic molecules the standard DFT+LDA approach\
leads to the orders-of-magnitude discrepancy between the measured and
calculated currents and to incorrect predictions of the conducting (instead
of experimentally observed insulated) phase\cite%
{Evers,Toher,Koentopp,Palacios,Muralidharan}. There have been attempts to
explain these discrepancies by insufficient modelling, such as atomistic
structures or contact coupling. Several recent studies however, attributed
this discrepancy to more fundamental reasons, identifying the lack of the
derivative discontinuity in LDA as a major source of error in the DFT-based
transport calculations\cite{Toher,Koentopp}. For example, Toher \textit{et
al.}\cite{Toher} argue that LDA approximation is not suitable for transport
calculations for the case of weak coupling. They propose a simple corrective
scheme based on the removal of the atomic self-interaction. Their corrective
scheme restores an agrement with the experiment, opening a conduction gap of
the \textit{I-V} characteristics of a molecular junction instead of a
metallic behavior following from the standard DFT+LDA approach. (It is
interesting to note that similar self-interaction corrections practically do
not effect the conductance and electronic density for the case of
strongly-coupled systems\cite{Toher}). Alternative corrective schemes
replacing the Kohn-Sham data for the weakly coupled region with their
counterparts obtained from a Hartree-Fock analysis (taking care of the
self-interaction problem) are suggested by Evers \textit{et al.}\cite{Evers}
and Palacios\cite{Palacios}. Note that various approaches to the description
of quantum transport for the case of weakly coupled systems (accounting for
the charge quantization and thus eliminating the self-interaction errors)
were discussed in Refs. \onlinecite{Indlekofer,Wacker,Muralidharan}.

We argue here that a similar problem related to the derivative discontinuity
may be the reason why the standard DFT+LDA approach fail to describe the
observed 0.7-anomaly in the QPC. Indeed, the formation of a spin-polarized
charge droplet predicted within the DFT+LDA approach implies that electrons
are trapped in weakly coupled quasi-bound states in the center of the QPC.
As mentioned above, in the case of weak coupling the lack of the derivative
discontinuity in the LDA approximation causes the orders-of-magnitude
discrepancies between the theory and experiment for the molecular systems.
Thus, it would be reasonable to expect that due to the same reason the LDA
approximation is not suitable for the case of the QPC structure as well.
Because the corrective schemes accounting for the derivative discontinuity
are shown to strongly affect the electron density and the energy levels in
the system, and because of the apparent failure of the standard DFT+LDA
approach to reproduce the 0.7 anomaly, we conclude that the formation of the
magnetic impurities in the QPC might be an artefact of the LDA due to the
lack of the derivative discontinuity related to the spurious
self-interaction.

Based on the above discussion we conclude that an accurate
description of the QPC structure might require approaches that go
beyond the standard DFT+LDA scheme and account for the derivative
discontinuity in the $V_{xc}$ or utilize similar corrective
schemes eliminating the self-interaction errors of the DFT+LDA. It
is not clear at the moment whether the recipes for the accounting
of the derivative discontinuity and eliminating the
self-interaction correction developed for the molecular junctions\cite%
{Toher,Evers,Palacios} can be adapted for the QPC structure. Such the
corrective schemes are remained to be implemented and it remains to be seen
whether this can bring the calculated conductance to the closer agreement
with the experiment.

As an indirect support of the above arguments we notice that similar spin
DFT conductance calculations (within the same LDA approach and the same
parameterization of $V_{xc}$)\cite{Marcus_paper} reproduce \textit{%
quantitatively} the measured spin-resolved magnetoconductance of the QPCs in
the integer quantum Hall regime\cite{Radu}. In this case the edge state
regime is reached such that the transport through the QPC correspond to the
strong coupling regime.

Finally, a comment is in order concerning the Kondo physics. The Kondo
effect was suggested by Cronenwett \textit{et al.}\cite{Cronenwett} as a
possible source of the 0.7 anomaly; at the same time, the experimental
studies of Graham\cite{Graham} Rokhinson\cite{Rokhinson} seem to rule out
this interpretation. We stress that the present mean-field approach based on
the standard DFT+LDA formulation is not able to address this effect.
Moreover, we do not believe that the prediction of the magnetic impurity
formation within the DFT+LDA approach can support or rule out the Kondo
physics in the QPC. Indeed, the Kondo enhanced conductance channels (if they
exist) would change the electron density inside the QPC (beyond that one
predicted by the DFT). Thus, the DFT+LDA predictions for the equilibrium
electron density in the QPC should be corrected in a self-consistent way to
account for this additional density, and it is not obvious whether the
spin-polarized quasi-bound states predicted in the framework of the DFT+LDA
would survive this correction.

\section{Conclusion}

We have developed an approach for full quantum mechanical transport
conductance calculations in open lateral split-gate structures that starts
from the lithographical layout of the device and is free from
phenomenological parameters like coupling strengths, charging constants etc.
The whole structure, including the semi-infinitive leads, is treated on the
same footing (i.e. the electron-electron interaction is accounted for in
both the leads as well as in the central device region). The
electron-electron interaction and spin effects are included within the spin
density functional theory in the local spin density approximation, and the
conductance calculated on the basis of the standard Green's function
technique.

The developed method was applied to calculate the spin-resolved conductance
through a QPC. Close to the pinch off a spin-degeneracy of the spin-up and
spin-down conductance channels is lifted and the total conductance shows a
broad feature peaked at $0.5\times 2e^{2}/h$ (corresponding to a complete
suppression of one of the spin channels). A similar feature is also present
in the range of the gate voltages where a second step in the conductance
develops. The lifting of the spin-degeneracy and the suppression of one of
the spin channels are the generic features of all studied QPCs (both very
short and very long; $40\lesssim l\lesssim 500$ nm). The calculated
conductance also shows a hysteresis for forward- and backward sweeps of the
magnetic fields. These features in the conductance are related to the
formation of weakly coupled (quasi-bound) states in the constriction of the
QPC (predicted in previous DFT-based studies\cite{Meir_2003,Meir_Nature}).
\textit{A comparison of the obtained results with the experimental data
shows however, that while the spin-DFT based \textquotedblleft
first-principle\textquotedblright\ calculations predict the spin
polarization in the QPC structure, the calculated conductance clearly does
not reproduce the 0.7 anomaly observed in almost all QPCs of various
geometries.}

In order to understand why the calculated conductance fails to reproduce the
0.7 anomaly, we critically examine the major features of the DFT-based
conductance calculations. We suggest that inability of the standard DFT+LDA
approximation to reproduce the 0.7 anomaly might be related to infamous
derivative discontinuity problem of the DFT leading to spurious
self-interaction errors not corrected in the standard LDA\cite%
{Giuliani_Vignale}. This problem has been recently recognized in similar
DFT-based transport calculations for the molecular junctions (showing
orders-of-magnitude discrepancies with the experiment) where it has been
demonstrated that the LDA approximation is not suitable for transport
calculations for the case of weak coupling\cite{Toher,Koentopp}. We thus
conclude that the formation of the weakly coupled magnetic impurities in the
QPC might be an artefact of the LDA due to the lack of the derivative
discontinuity and related spurious self-interaction. Our results suggest
that an accurate description of the QPC structure requires approaches that
go beyond the standard DFT+LDA scheme and that account for the derivative
discontinuity in the $V_{xc}$ or utilize similar corrective schemes
eliminating the self-interaction errors of the DFT+LDA.

\begin{acknowledgments}
S. I. acknowledges financial support from the Swedish Institute. Numerical
calculations were performed in part using the facilities of the National
Supercomputer Center, Link\"{o}ping, Sweden.
\end{acknowledgments}

\end{document}